\documentclass [12pt]{article}

\usepackage[T2A]{fontenc}

\usepackage[cp1251]{inputenc}
\usepackage[english]{babel}
\usepackage{amssymb}

\usepackage{epsfig}

\usepackage{subcaption}

\newcommand{\be}{\begin{equation}}
\newcommand{\ee}{\end{equation}}
\newcommand{\bea}{\begin{eqnarray}}
\newcommand{\eea}{\end{eqnarray}}

\def\p{\partial}
\def\pslash{\p\raise.3ex \hbox{\kern-.5em /}}
\def\delslash{\nabla\raise.3ex \hbox{\kern-.7em /}}

\begin{document}
\begin{center}
\Large{ \textbf{  Towards  the  detecting  of pseudo-Hermitian
anomalies for negative square masses neutrinos  in intensive
magnetic fields }}
\end{center}
\vskip 0.2cm
\begin{center} \Large{V.N.Rodionov}
\footnote{Plekhanov Russian University of Economics, Moscow,
Russia, \em E-mail: rodyvn@mail.ru}
\end{center}

\begin{center}

\abstract{One of the primary goals of contemporary physics of
neutrinos after discovery of  their masses  become the
investigation of their  electromagnetic properties. This is a
necessary step for creation  of new physics beyond the Standard
Model (SM),   which no longer  can claim the role theory
explaining everything phenomenon of  the Universe. On this   it
should draw attention because SM remains consistent local  scheme
for any value of the  masses of  particles $0\leq m <\infty $. Now
the masses of elementary particles  can exceed even  Planck's mass
 $m_{Planck}\simeq 10^{19}GeV$, which is the largest scale mass
 in the Universe. For solving this problem of studying of
 electromagnetic interactions of neutrino  we suggest use  the
 methods of  relativistic quantum theory with the limiting
 mass $m\leq M$.  The restriction of mass spectrum of fermions can
 be obtained in the  frame  of non-Hermitian (pseudo-Hermitian)
 fermion systems having the  direct application to the neutrino physics.
 The systems of the similar type  include so-called parity-time-symmetric
 (${\cal PTS}$) -  models, which already are used in various fields
 of modern  physics and, in particular for experimental investigation
 of ${\cal PTS}$-optics. We also hope that experimental problem
 neutrinos  with the negative mass squared could be also solved
 on the basis of ${\cal PTS}$-notions  taking into account
 the characteristics of neutrinos  of new type ("exotic neutrino")
 emerging in the theory with the Maximal Mass.  }
\end{center}

  {\em PACS    numbers:  02.30.Jr, 03.65.-w, 03.65.Ge,
12.10.-g, 12.20.-m}

\vskip 0.5cm

\section{Introduction}

It is well known the Nobel Prize in Physics was awarded in 2015
jointly to Takaaki Kajita and Arthur B. McDonald "for the
discovery of neutrino oscillations, which shows that neutrinos
have mass". This discovery has completely changed our
understanding of the innermost properties of matter and showed
that SM cannot be the comprehensive theory of the fundamental
constituents of the Universe. However obviously, that of past
successes of the SM is so high that new models which is designed
for modifying SM, must contain practically all basic principles
lying in the basis of already existing theory. In this connection
we should to note that main extension SM may be  connected  with a
generalization of  notion   Hermiticity.

 Indeed it is known  that one of the fundamental postulates of
quantum theory is the requirement of Hermiticity of physical
parameters. This condition not only guarantees the reality of the
eigenvalues of Hamilton operators, but also implies the
preservation in time of the probabilities of the considered
quantum processes. However, as it was shown \cite{ben},
Hermiticity is a sufficient but it is not a necessary condition.
It turned out that among non-Hermitian Hamiltonians it is possible
to allocate a wide range pseudo-Hermitian Hamiltonians with
providing the real energy spectrum and  the development of systems
over time with preserving unitarity.

  Now it is well-known fact, that the reality of the
spectrum in models with a non-Hermitian Hamiltonian is a
consequence of $\cal PT$-invariance of the theory, i.e. a
combination of spatial and temporary parity of the total
Hamiltonian: $[H,{\cal PT}]\psi =0$. When the $\cal PT$ symmetry
is unbroken, the spectrum of the quantum theory will be  real.
This surprising results explain the growing interest in this
problem which was initiated by Bender and Boettcher's observation
\cite{ben}. For the past a few years has been studied a lot of new
non-Hermitian $\cal PT$-invariant systems (see, for example
\cite{Makris1}-\cite{RodKr1}).

The algebraic non-Hermitian ${\cal PTS}$ $\gamma_5$-extension of
the Dirac equation was first studied in \cite{ft12} and further
was developed  in our works \cite{ROD1}-\cite{RodKr1}. But in the
geometrical approach to the construction of  Quantum Field Theory
(QFT) with fundamental mass which was developed by V.G.Kadyshevsky
and his colleagues (\cite{Kad1}-\cite{Max}), already has contained
equations for motion of fermion with $\gamma_5$-mass extension.
Besides   in the geometrical QFT we can watch the emergence of new
particles, which were named "exotic particles" \cite{Kad1}.  At
first their the emergence, really  considered the prerogative of
the geometric approach. However now it is clear, that appearance
of such non-ordinary particles  is the consequence of using of
non-Hermitian models  \cite{Rod1}-\cite{RodKr1}.

This type of Hamiltonians includes the so-called ${\cal
PTS}$-models, which already used in various fields of modern
physics. The most developed in this respect are such which used in
the field of ${\cal PTS}$-optics, where for several years
conducted not only theoretical, but also experimental studies
\cite{Makris1},\cite{Makris2}. Although ordinary Hermite
interpretation does not disputed ${\cal PTS}$-approach may open up
new non-trivial properties early well-studied objects. Thus,
development of non-Hermitian approach becomes a very fruitful
environment for the creation of new physics beyond the SM
\cite{ROD1},\cite{ROD2}.

In this connection we want to draw  attention that candidate for
the new quantum theory of fermions  is the modified model with
pseudo-Hermitian $\gamma_5$-mass extension of the ordinary quantum
Dirac theory. This model can be constructed  using  the simple
generalization Dirac's approach to obtaining his  known equations
of motion fermions.  The Hamiltonian form of this new equations
explicitly shows that these Hamiltonian are non-Hermitian, but
$\cal PT$-invariant and pseudo-Hermitian. This condition not only
guarantees the reality of the eigenvalues of Hamilton operators,
but also implies the preservation in time of the probabilities of
the considered quantum processes \cite{RodKr1}.

In last decade, many cosmological observations suggest the
existence of dark matter and dark energy and they offer to
consider them as key components of the content of the Universe
(see for example \cite{dong1},\cite{dong2}). The goal of detecting
the nature of dark matter triggered the development a number of
new, highly sensitive detectors that can catch extremely weak
interactions between dark matter and objects belonging to SM
\cite{01}. Nevertheless, in the literature already exist
indications of the existence of the candidates to the structure of
dark matter, which is required verification. In particular, there
are particles, description of which non corresponds Hermitian
theories, but arise in the models with preserving  ${\cal PTS}$
\cite{Rod1}-\cite{RodKr1}.
% It is shown that the models with
%$\gamma_5$-mass extension not only let to obtain modified Dirac
%theory but also have the descriptions of any new fermions  (exotic
%neutrinos), which can arise in this model .

 In present paper, we develop the algebraic ${\cal PTS}$-model
 in the intensive magnetic field,
in the frame of which early we have obtained the exact solutions
for the real eigenvalues of the energy of charged and neutral
fermions \cite{ROD1},\cite{ROD2}.       It is very important that
there are occur  violation of $\cal PT $-symmetry, depending on
the intensity of the magnetic field and the orientation of fermion
spin.   According to last investigations in this algebraic model
also \emph{may   be restriction of mass spectrum of fermions}
\cite{ROD1}-\cite{RodKr1}.
  This is limit of mass has far-reaching
consequences in the form of the appearance in theory of new
particles, no existing in the SM. The main difference of these
particles reduced to that they do not obey to the ordinary Dirac
equations. For them do not known all properties, but they have
equal mass with its partners from ordinary particles. The most
perspective for investigation are "exotic neutrinos". At present
one can speak about a wide spectrum of different type neutrinos.

It is enough to remember that there are  sterile and active
neutrinos. Moreover sterile neutrinos may mix with active
neutrinos via a matrix mass. Existence masses of neutrino  and
mixing implies that neutrinos  have magnetic moments. In last time
one can often meet with  an overviews   of electromagnetic
properties neutrino, (see, for example, \cite{Ternov1}). But as it
was noted up in this paper " now there is no positive experimental
indication in favor existence electromagnetic properties of
neutrinos".  With it really is hard not to agree because  the
interactions of ordinary neutrinos with the electromagnetic fields
are extremely weak.  However if one to suggest using the "exotic
neutrinos" the interaction with magnetic field  may be really
significantly increased thanks to the coefficient which be equal
to the ratio of Maximal Mass and mass of   neutrinos
$k=M/m_{\nu}$\cite{ROD1},\cite{ROD2}. Such experiments in our
opinion may  be very fruitful for creation  the new physics beyond
the SM. Perhaps that  this effects  indeed can be observed in
terrestrial experiments.  In the papers \cite{ROD1},\cite{ROD2}
the  energy spectra of the fermions was obtained by us as exact
analytical solutions of the modified Dirac equation  which taken
into account with the interaction of anomalous magnetic moments
(AMM) of fermions with intense magnetic fields.

Note also that intense magnetic fields exist in the series of
space objects. In particular, magnetic field intensity of the
order of $10^{12}\div 10^{13}$ Gauss observed near and inside of
the pulsars. Here also may be noted recent discovered objects
which are known as sources of soft repeating gamma ray bursts and
anomalous X-ray pulsars. Magneto-rotational model was proposed for
their description and they were named the magnetars. It is
suggested that for such objects, the attainable magnetic fields
with a strength up to $10^{15}$ Gauss. This is very important what
fraction of magnetars in the General population of neutron stars
can reach 10\%\cite{4}. In this connection,  we note that
processes involving neutrinos in the presence of such strong
magnetic fields can render a significant impact on processes that
can shape evolution of the astrophysical objects. It can be
especially noticeable if in a stream of neutrino may be a certain
percentage of particles with non-Hermitian characteristics.

On the other hand in 1965  M.A.Markov \cite{Mar} has proposed
hypothesis   according to which the mass spectrum of particles
should be limited by  M.Planck mass $m_{Planck} =\sqrt{\hbar
c/G}\approx 10^{19} GeV $. The particles with the limiting mass
\be\label{Markov}m \leq m_{Planck}\ee were named by the author the
Maximons. However, condition (\ref{Markov}) initially was purely
phenomenological and until recently it has seemed that this
restriction  can be applied without connection with SM.  And
really SM is irreproachable scheme for value of mass from zero
till infinity. But in the current situation, however, more and
more data are accumulated which bear witness in favor of the
necessity of revising notion of Hermiticity. In particular, this
is confirmed by abundant evidence that "dark matter", really
exists and absorbs a substantial part of the energy density in the
Universe.

A new radical approach was offered by V. G. Kadyshevsky and his
colleagues \cite{Kad1}-\cite{Max}, in which the Markov's idea of
the existence of a Maximal Mass used as new fundamental principle
construction of QFT. This principle refutes the affirmation that
mass of the elementary particle can have a value in the interval
$0 \leq m < \infty$. In the geometrical theory the condition
finiteness of the mass spectrum is postulated in the form
\be\label{M} m \leq {\cal M},\ee where the Maximal Mass parameter
${\cal M}$, was named by the the \emph{fundamental mass}. This
physical parameter is a \emph{ new physical constant} along at the
speed of light and Planck's constant. The value of ${\cal M}$ is
considered as a curvature radius of a five dimensional hyperboloid
whose surface is a realization of the curved momentum 4-space --
the anti de Sitter space. Objects with a mass larger than ${\cal
M}$ cannot be regarded as elementary particles because no local
fields that correspond to them. For a free particle, condition
(\ref{M}) accomplished automatically on surface of a five
dimensional hyperboloid. In the approximation ${\cal M} \gg m$ the
anti de Sitter geometry goes over into the Minkowski geometry in
the four dimensional pseudo-Euclidean momentum space ("flat
limit")\cite{Kad1}.

\section{ Modified model for the study of non-Hermitian mass parameters}

 Let us now consider the modified Dirac equations for
free massive particles using the ${\gamma_5}$-factorization of the
ordinary Klein-Gordon operator. In this case we will make similar
actions as for known Dirac procedure. He particularly  has wrote
"...need get something like a square root from the operator of
Klein-Gordon" \cite{Dirac1}. And really if we shall not be
restricted to only Hermitian operators then we can represent the
Klein-Gordon operator in the form of a product of two commuting
matrix operators with $\gamma_5$-extension of mass:

\be\label{D2} \Big({\partial_\mu}^2 +m^2\Big)=
\Big(i\partial_\mu\gamma^{\mu}-m_1-\gamma_5 m_2 \Big)
\Big(-i\partial_\mu\gamma^{\mu}-m_1+\gamma_5 m_2 \Big), \ee where
 the physical mass of particles $m$ is expressed through the
 new parameters $m_1$ and
$m_2$ \be \label{012} m^2={m_1}^2- {m_2}^2. \ee

For  the function which  will obey to the equation of Klein-Gordon
\be\label{KG} \Big({\partial_\mu}^2
+m^2\Big)\widetilde{\psi}(x,t)=0 \ee one can demand that it also
satisfies to one of the equations of the first order
\be\label{ModDir} \Big(i\partial_\mu\gamma^{\mu}-m_1-\gamma_5 m_2
\Big)\widetilde{\psi}(x,t)
=0;\,\,\,\Big(-i\partial_\mu\gamma^{\mu}-m_1+\gamma_5 m_2 \Big)
\widetilde{\psi}(x,t)=0 \ee

%\begin{figure}[h]
%\vspace{-0.2cm} \centering
%\includegraphics[angle=0, scale=0.5]{Fig6-1.eps}
%\caption{Dependence of $y(\alpha)= m/M $ on the parameter
%$\alpha$} \vspace{-0.1cm}\label{Fig.6-0}
%\end{figure}

Equations (\ref{ModDir}) of course, are less common than
(\ref{KG}), and although every solution of one of the equations
(\ref{ModDir}) satisfies to (\ref{KG}), reverse approval has not
designated. It is also obvious that the Hamiltonians, associated
with the equations (\ref{ModDir}), are non-Hermitian
(pseudo-Hermitian), because in them the $\gamma_5$-dependent mass
components appear ($H\neq H^{+}$):

  \be\label{H0} H =\overrightarrow{\alpha} \textbf{p}+ \beta(m_1
+\gamma_5 m_2)=\overrightarrow{\alpha} \textbf{p}+ \beta m
e^{\gamma_5 \alpha} \ee and \be\label{H+} H^+
=\overrightarrow{\alpha }\textbf{p}+ \beta(m_1 -\gamma_5
m_2)=\overrightarrow{\alpha} \textbf{p}+ \beta m e^{-\gamma_5
\alpha}.\ee Here  matrices $\alpha_i=\gamma_0\cdot\gamma_i$,
$\beta=\gamma_0$, $\gamma_5=-i\gamma_0\gamma_1\gamma_2\gamma_3$
and introduced identical replacement of parameters
\be\label{alpha}\sinh(\alpha)=m_2/m;
\,\,\,\,\cosh(\alpha)=m_1/m,\ee where parameter $\alpha$ varies
from zero to infinity.

Additionally, an alternative formalism for regarding the systems
defined by non-Hermitian Hamiltonians  is also known, according to
which the spectrum reality for a non-Hermitian system occurs owing
to the so-called pseudo-Hermitian properties of a Hamiltonian. A
Hamiltonian is called pseudo-Hermitian if it satisfies the
condition
      $$ H^+ =\eta_0 H {\eta_0}^{-1},     $$
where $\eta_0$ is a linear Hermitian operator. From (\ref{H0}) and
(\ref{H+}) we can find
$$
  \eta_0= e^{\gamma_5\alpha}.
$$

It is easy to see from (\ref{012}) that the  mass $m$, appearing
in the equation (\ref{KG}) is real, when the inequality \be
\label{e210} {m_1}^2\geq {m_2}^2,\ee is accomplished. However for
variable $\alpha$ which is identical for definitions
$m_1,\,\,\,m_2$ this  condition is automatically accomplished in
all region $0\leq\alpha<\infty.$

However this area contains descriptions not only pseudo-Hermitian
fermions, which in a result of the Hermitian  transition ($m_2
\rightarrow 0, \,\, m_1\rightarrow m$)  coincide with the ordinary
particles, besides also there are the second region. In this case
fermions don't  subordinate to the ordinary Dirac equations and
for them the Hermitian limit is absent. This  is easy to see that
if we shall be used for new restrictions of mass parameters the
simple expression. Really, if we take into account inequality
between arithmetic and geometrical averages of two positive
numbers we have

\be\label{m2M} m^2 + {m_2}^2 \geq 2\sqrt{m^2\,{m_2}^2 }.\ee

\begin{figure}[p]
\begin{minipage}[c]{12cm}
\begin{minipage}[c]{5cm}
\centering
\includegraphics[height=4cm]{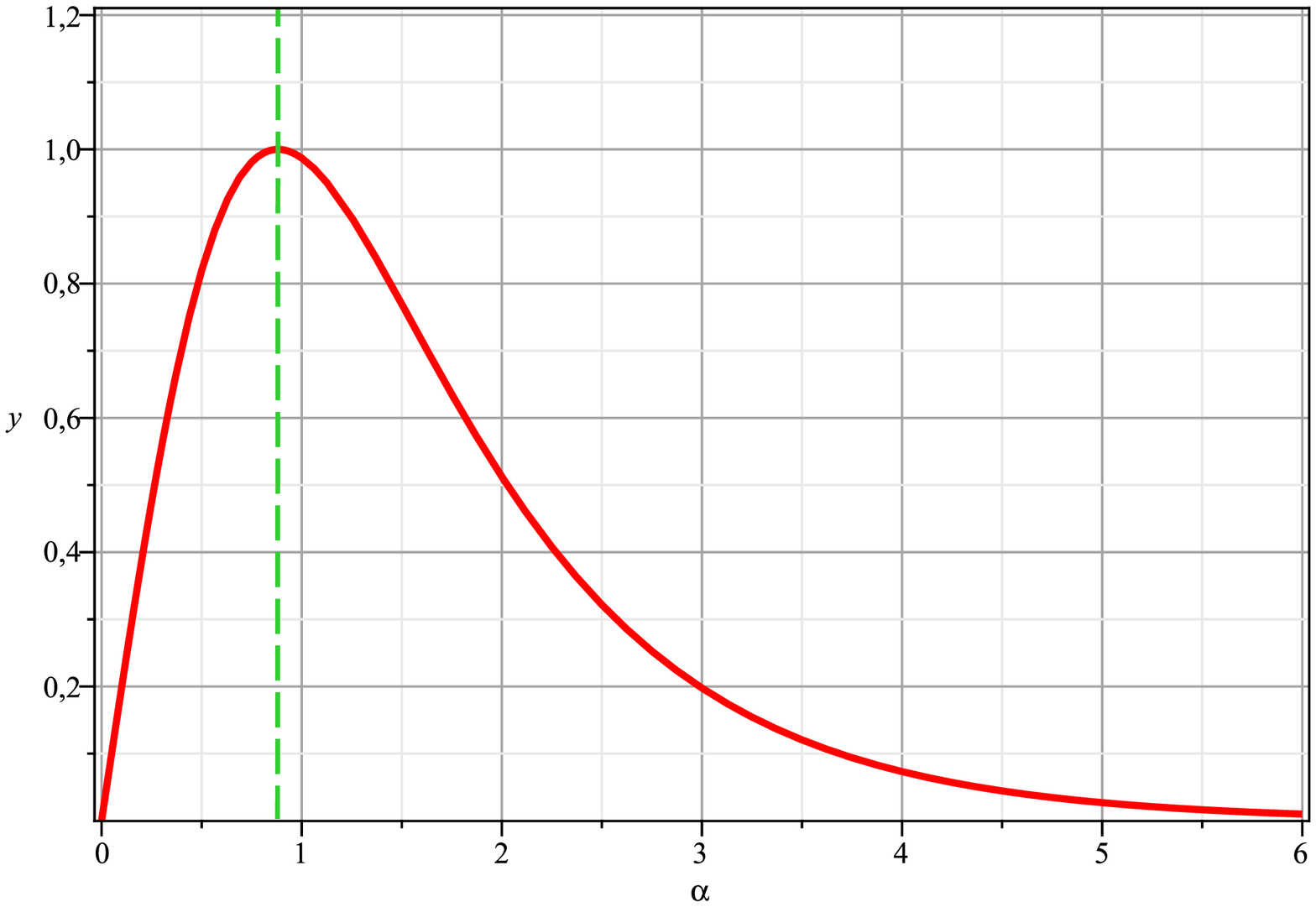}
\subcaption{}\label{Figalpha}
%Dependence of normalized values $m^{-}_1 /M; m^{-}_2 /M$
%for ordinary particles (downstairs)  and $m^{+}_1/M;  m^{+}_2 /M$
%for exotic particles (upstairs)  on the parameter $x=m/M$.
%} \label{Fig5-1}
\end{minipage}
\hfill
\begin{minipage}[c]{5cm}
\centering
\includegraphics[height=4cm]{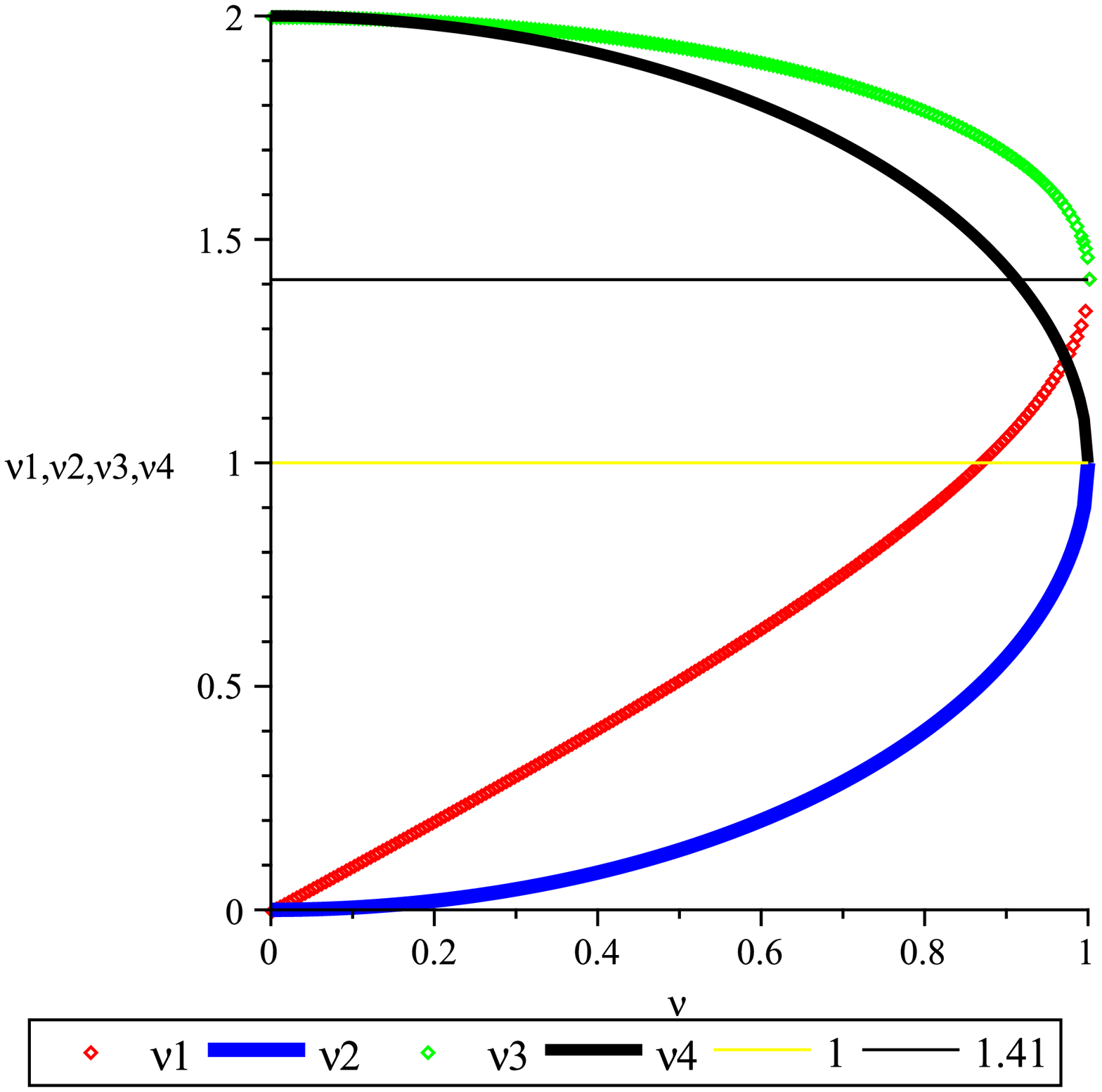}
\subcaption{}\label{Fig2}
%The different regions of undisturbed PT symmetry for
%ordinary and exotic particles
%} \label{Fig5-2}
\end{minipage}
\end{minipage}
\begin{minipage}[c]{12cm}
\begin{minipage}[c]{5cm}
 \centering
\includegraphics[height=4cm]{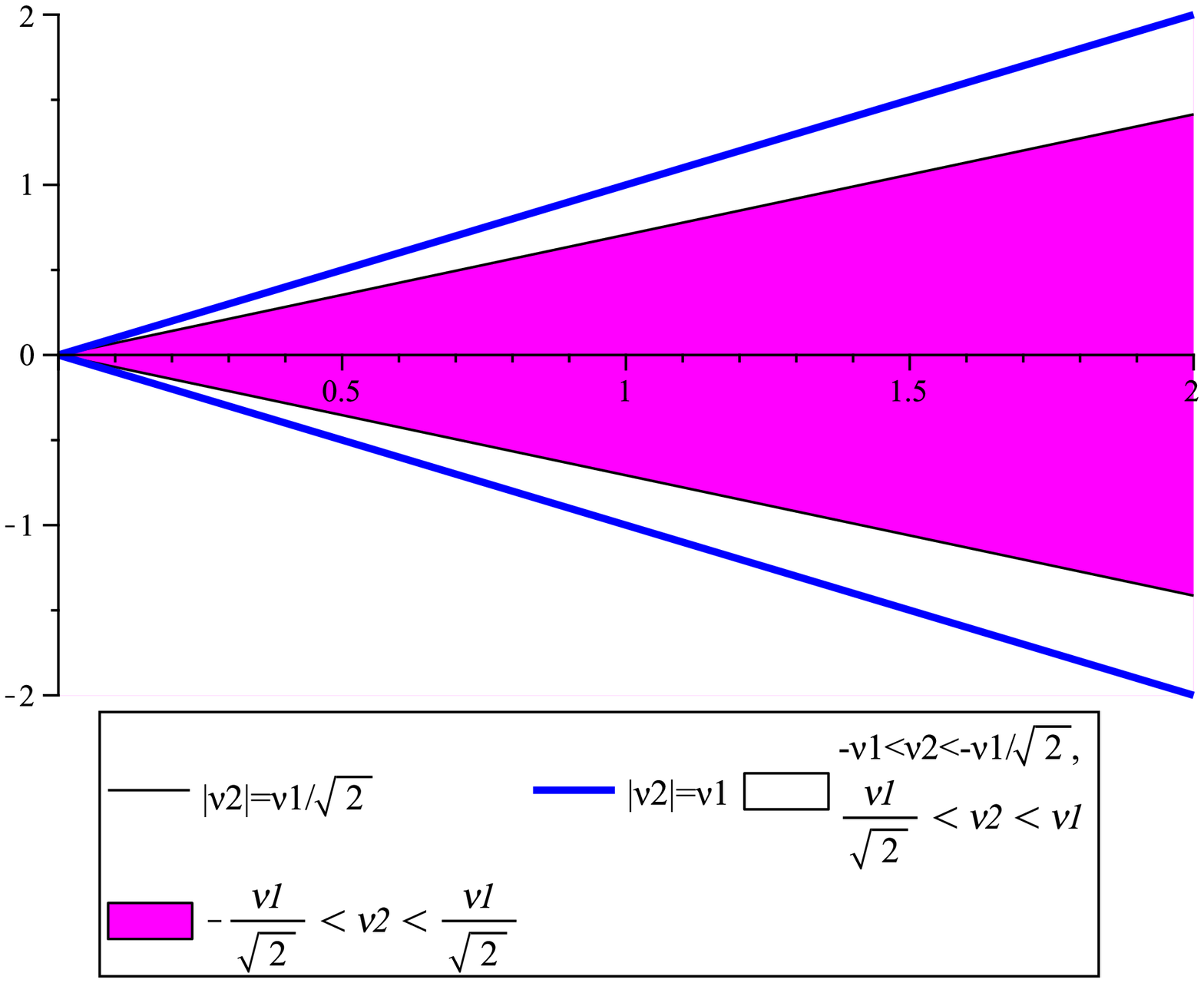}
\subcaption{}\label{Fig3}
%Dependence of of normalized values $E(-1,0, p_\bot,
%\Delta\mu H = 0.1 )/m$ on the parameter $x=m/M$ for the cases
%$p\bot=0,1,2,3,4\,\, and\,\,\Delta\mu H = 0.1.$
%} \label{Fig5-3}
\end{minipage}
\hfill
\begin{minipage}[c]{5cm}
 \centering
\includegraphics[height=4cm]{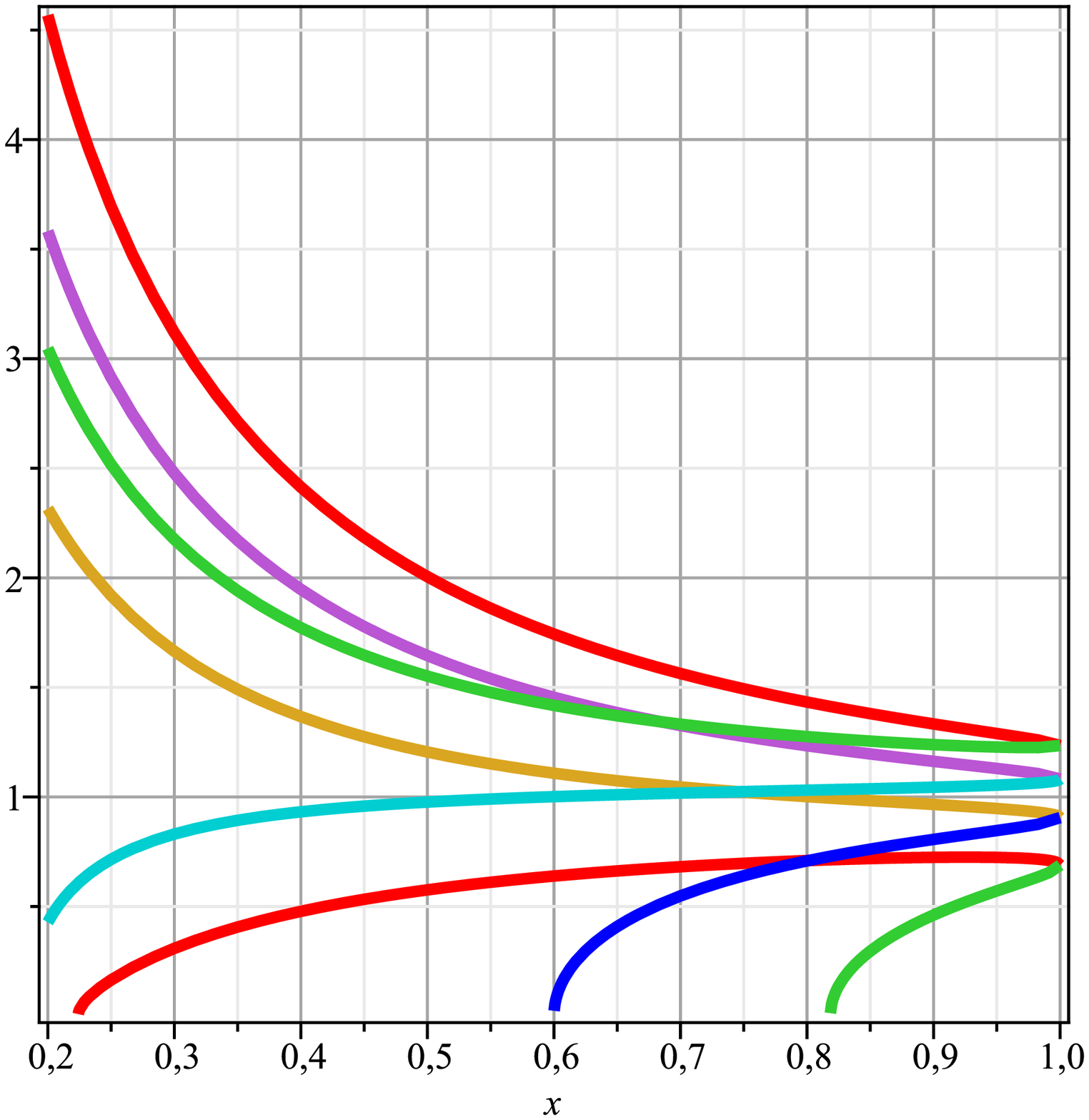}
\subcaption{}\label{Fig8}
%Dependence of $E(-1,p_3, p_\bot, \Delta\mu H = 0.1 )/M$
%on the longitudinal components of momentum $x=p_3/M$ for the
%different values of transverse momentum components $p_\bot/m$:
%$p\bot=0,1,2,3,4\,\, and\,\,\Delta\mu H = 0.1.$
%} \label{Fig5-4 }
\end{minipage}
\end{minipage}
\begin{minipage}[c]{12cm}
\begin{minipage}[c]{5cm}
 \centering
\includegraphics[height=4cm]{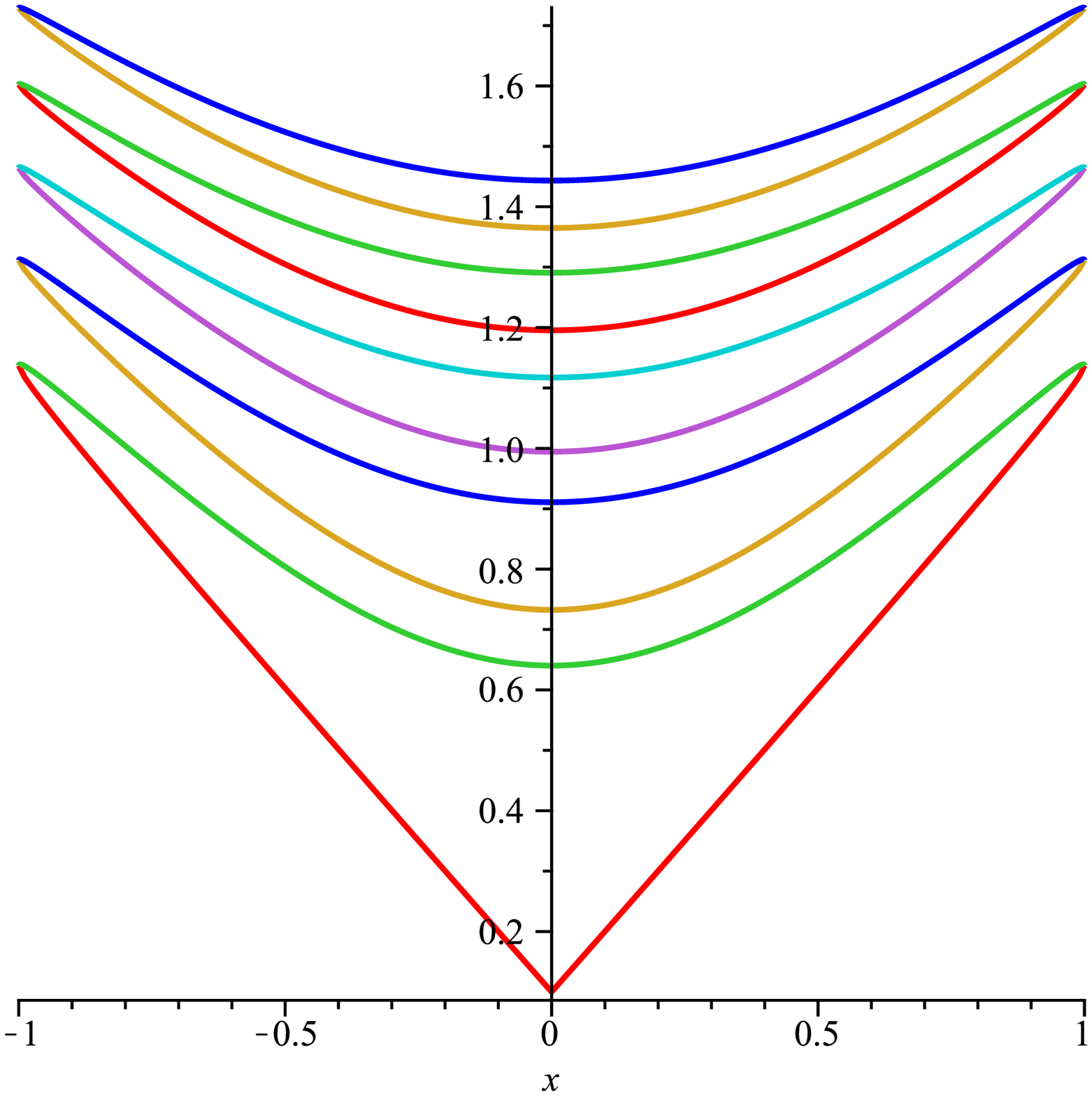}
\subcaption{}\label{Fig1p1}
%Dependence of $E(-1,0, p_\bot, \Delta\mu H = 0.1 )/m$ on
%the parameter $x=m/M$ for the different values of transverse
%momentum components $p_\bot/m$: $p\bot=0,1,2,3,4\,\,
%and\,\,\Delta\mu H = 0.1.$
%} \label{Fig5-5 }
\end{minipage}
\hfill
\begin{minipage}[c]{5cm}
 \centering
\includegraphics[height=4cm]{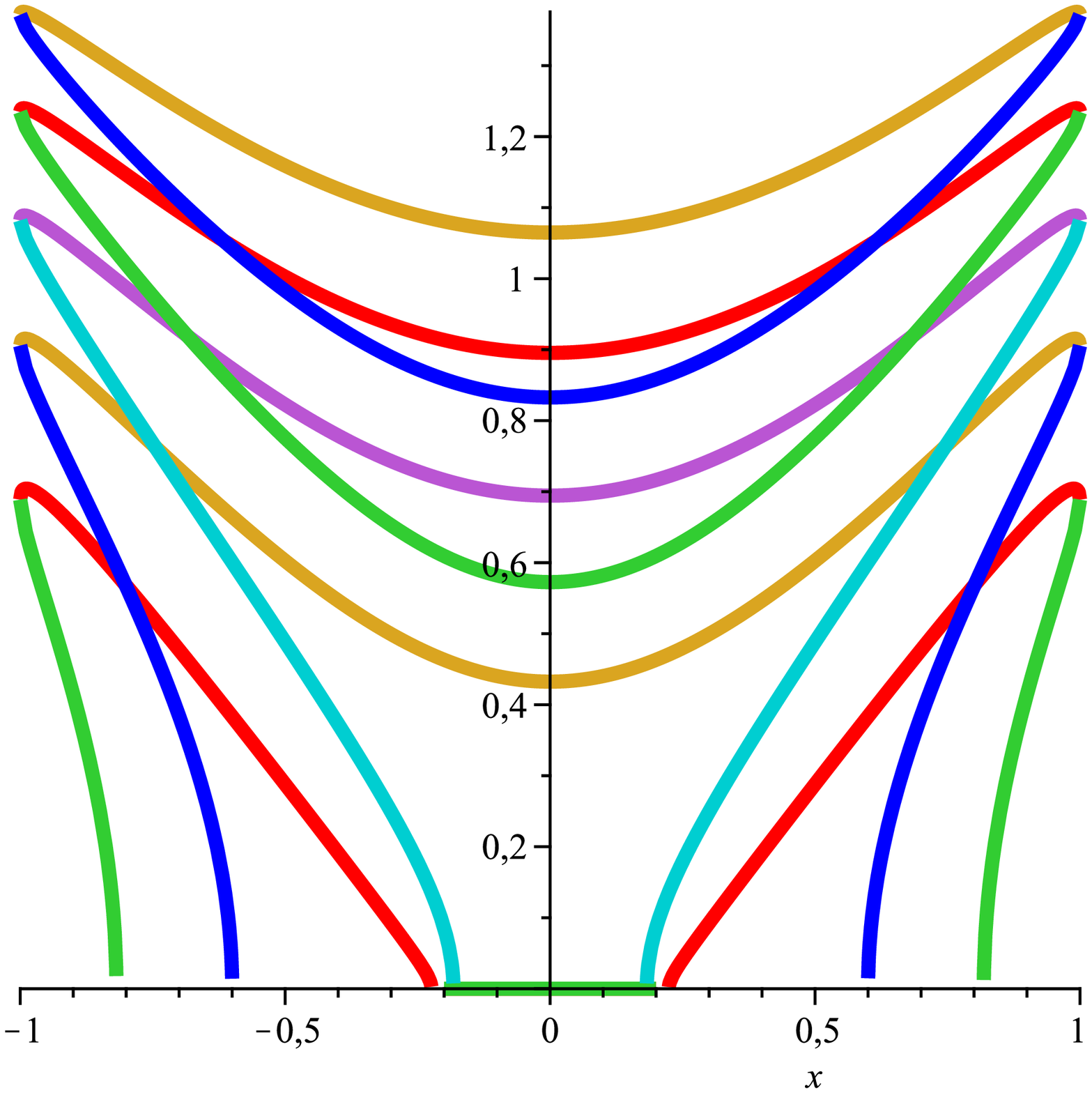}
\subcaption{}\label{Fig5-1}
%Dependence square of the energy  particle in the magnetic
%field $ E^2(\pm1,p_3, p_\bot, \Delta\mu H = 0.1 )$ on the
%longitudinal components of momentum $x=p_3/M$ on the direction of
%the magnetic field $x=p_3 /M$ for different spin orientation and
%different values of transverse momentum components $p_\bot$.
%} \label{Fig5-6}
\end{minipage}
\end{minipage}
\begin{minipage}[c]{12cm}
\begin{minipage}[c]{5cm}
 \centering
\includegraphics[height=4cm]{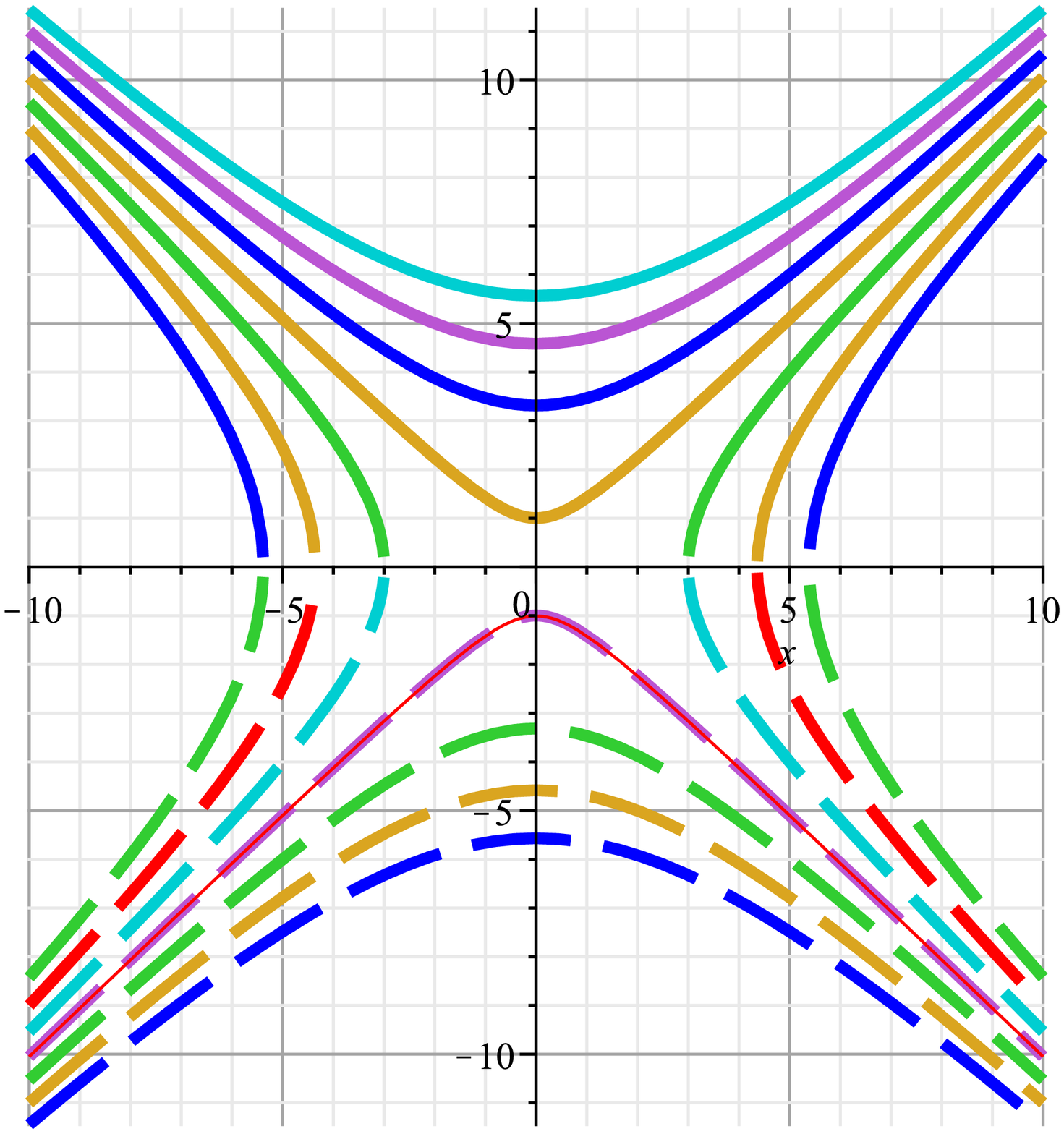}
\subcaption{}\label{Fig7}
%Dependence of $E(-1,0, p_\bot, \Delta\mu H = 0.1 )/m$ on
%the parameter $x=m/M$ for the different values of transverse
%momentum components $p_\bot/m$: $p\bot=0,1,2,3,4\,\,
%and\,\,\Delta\mu H = 0.1.$
%} \label{Fig5-5 }
\end{minipage}
\hfill
\begin{minipage}[c]{5cm}
 \centering
\includegraphics[height=4cm]{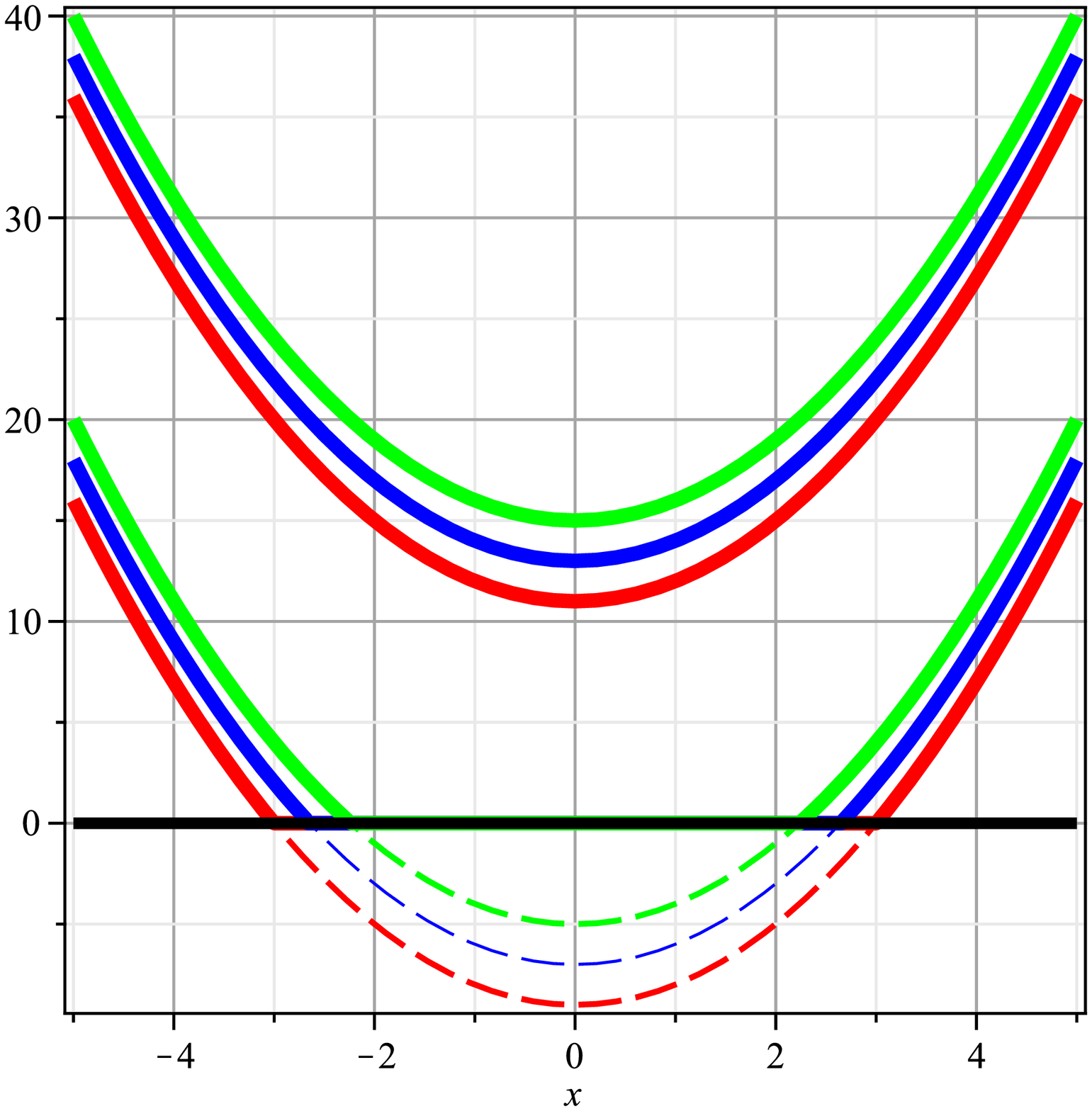}
\subcaption{}\label{Fig1111}
%Dependence square of the energy  particle in the magnetic
%field $ E^2(\pm1,p_3, p_\bot, \Delta\mu H = 0.1 )$ on the
%longitudinal components of momentum $x=p_3/M$ on the direction of
%the magnetic field $x=p_3 /M$ for different spin orientation and
%different values of transverse momentum components $p_\bot$.
%} \label{Fig5-6}
\end{minipage}
\end{minipage}
\caption{These  figure give possibility to verify that restriction
in  mass spectrum  of fermions $m\leq M$ obtained in the framework
of non-Hermitian relativistic  quantum theory  leads to appearance
of new particles -- the exotic fermions (the plots (a)-(d)).
Besides these  a number of illustrations (the plots (e)-(h))  bear
witness that  discovery of exotic neutrinos can indeed   open
approach to creation of new physics beyond the Standard Model.}
\end{figure}

Sign of equality takes place when $m = m_2$ and this mass
$m=m_2=M$ has Maximal allowable value for  particle mass. Besides
of from (\ref{012})  we can also see that  in this point
$m_1=\sqrt{2}M$. Thus value $M$ can be considered as Maximal
fermion Mass  in pseudo-Hermitian approach under fixed values
$m_1$ and $ m_2$ \be\label{mM} m\leq M.\ee In Markov's terminology
\cite{Mar} one can say that by such a way we defined the mass of
Maximon. It is easy to see this mass may be written in the
non-Hermitian form  \be \label{Mgamma5} m_{Maximon}=\sqrt{2}M +
\gamma_5 M.\ee   Using also expression (\ref{m2M}) one can also
obtain that in this task the maximal mass value is expressed by
the following way \be\label{MM}M={m_1}^2 /2 m_2. \ee
 At the
Fig.(\ref{Figalpha}) we can see explicit behavior of the reduced
mass distribution of $y(\alpha)=m/M$, depending on parameter
$\alpha.$ From this picture follows that the curve, corresponding
mass of the considered particles, which has a maximum  value of
$y(\alpha)= m/M = 1 $ in the point $\alpha_0 =0.881$  and  as
already noted correspond to Maximon. Till to this value we are
dealing with fermions, which have Hermitian limit when
$M\rightarrow \infty$ (or $m_2\rightarrow 0$). But after the value
$\alpha_0 =0.881$, where the maximal value of mass is achieved we
deal  with  decreasing mass of particles. However in this region
already no the possibility with the help of the limiting
transition to obtain Hermitian mass. Thus in this region particles
exist, which in principal differ from the particles of the SM. In
particular, massless description fermions may be represented in
the form:  $$ m_{massless}= 2M(1+\gamma_5), $$ where $M$ is
Maximal Mass of fermion spectrum. Thanks to the presence of a
Maximal value of mass parameter we can say about arising of  new
particles, which no exist in SM.

If the restriction of mass spectrum of elementary particles
doesn't exist in Nature then "exotic particles" can't arise. And
vice versa, if one can detect their presence it means that
limiting mass of fermions really exist.

It is very interesting that early such particles has been observed
in geometrical approach to the construction of QFT with
fundamental mass \cite{Kad1}-\cite{Max}. We believe that exact
solutions of modified Dirac-Pauli equations which were obtained by
us for the pseudo-Hermitian neutrinos can let valuable information
to detect the presence of such  "exotic particles". On this
indicates also huge increasing of the interaction exotic particles
with magnetic fields associated with unusual properties of "exotic
neutrinos" wich will be considered in the next section.

\section{Motion of Dirac particles, if their own magnetic moment is
different from the  Bohr's magneton}

In this section, we will want touch upon question about
description of the motion of   new fermions in the magnetic
fields, if their own magnetic moment is different from the  Bohr's
magneton. As it was shown by Schwinger \cite{Sc}  the equation of
Dirac particles in the external electromagnetic field $A^{ext}$
taking into account the radiative corrections may be represented
in the form: \be\label{A} \left({\cal P}\gamma -
m\right)\Psi(x)-\int{\cal M}(x,y|A^{ext})\Psi(y)dy=0, \ee where
${\cal M}(x,y|A^{ext})$ is the mass operator of the fermion in the
external field and ${\cal P}_\mu =p_\mu - {A^{ext}}_\mu$ . From
equation (\ref{A}) by means of expansion of the mass operator in a
series of according to $ eA^{ext}$ with precision not over then
linear field terms one can obtain the modified equation. This
equation preserves the relativistic covariance and consistent with
the phenomenological equation of Pauli obtained in his early
papers (see for example \cite{TKR}).

Now let us consider the model of massive fermions with
$\gamma_5$-extension of mass $m\rightarrow m_1+\gamma_5 m_2$
taking into account the interaction of their charges and AMM with
the electromagnetic field $F_{\mu\nu}$:

\be\label{Delta} \left( \gamma^\mu {\cal P}_\mu -
 m_1 -\gamma_5 m_2 -\frac{\Delta\mu}{2}\sigma^{\mu \nu}F_{\mu\nu}\right)\widetilde{\Psi}(x)=0,\ee
where $\Delta\mu = (\mu-\mu_0)= \mu_0(g-2)/2$. Here $\mu$ -
magnetic moment of a fermion, $g$ - fermion gyromagnetic factor,
$\mu_0=|e|/2m$ - the Bohr magneton,
$\sigma^{\mu\nu}=i/2(\gamma^\mu \gamma^\nu-\gamma^\nu
\gamma^\mu)$. Thus phenomenological constant $\Delta\mu $, which
was introduced by Pauli,  is part of the equation and gets the
interpretation with the point of view QFT.

The Hamiltonian form of (\ref{Delta}) in the homogenies magnetic
field is the following \be i\frac{\partial}{\partial t}
\widetilde{\Psi}(r,t)=H_{\Delta \mu}\widetilde{\Psi}(r,t),\ee
where \be\label{Delta1} H_{\Delta\mu} = \vec{\alpha}\vec{{\cal P}}
+ \beta(m_1 + \gamma_5 m_2) +
\Delta\mu\beta(\vec{\sigma}\textbf{H}).\ee

As it was noted earlier a feature of the model with $ \gamma_5
$-mass contribution is that it may contain another any
restrictions of mass parameters in addition to (\ref{e210}).
Indeed  for the physical mass $m$ may be constructed by infinite
number combinations of $ m_1 $ and $ m_2 $, satisfying to
(\ref{012}). However besides it need take into account the rules
of conformity of this parameters in the Hermitian limit $m_2
\rightarrow 0 $. Without this assumption the development of
Non-Hermitian models not may  be adequate. With this purpose one
can determine an additional mass scale $M$ which will depend on
$m_1$, $m_2$ (see (\ref{MM})). Indeed this parameter $M$ can put
an upper bound on the mass spectrum of particles. Thus fixing of
values $m$ and $M$ as experimentally significant and using
(\ref{012}) and (\ref{MM}) we can obtain expressions defining
dependence parameters $m_1$ and $m_2$ on them:
  \be \label{m1} {m_1}^{\mp}=
  \sqrt{2}M\sqrt{1\mp\sqrt{1-m^2/M^2}}\ee
  \be\label{m2} {m_2}^{\mp}=M(1\mp\sqrt{1-m^2/M^2})\ee

Using Fig.(\ref{Fig2}) we can see values of different branches of
${\nu_1}^\pm ={m_1}^\pm/M$ and ${\nu_2}^\pm={m_2}^\pm/M$  as a
function of normalized physical parameter $\nu=m/M$. Now  the
domain of the ${\cal PT}$-symmetry is defined by inequality $0\leq
\nu \leq 1$. For these values of $\nu $ the parameters $\nu_1$ and
$\nu_2$  correspond to   the modified Dirac equation with the
Maximal Mass  describing the propagation of particles with real
masses. But the lower branches ${\nu_1}^-$, ${\nu_2}^- $
correspond to ordinary particles and upper curves ${\nu_1}^+$,
${\nu_2}^+$ define the exotic  partners.

Difference of the ordinary non-Hermitian fermions and  exotic
particles can understand from the following reasoning. If we
accept that Maximal Mass in the fermion spectrum  is comparable
with Planck mass then using Fig.(\ref{Fig2}) it is easy to see,
that all known  particles   of SM   are accumulated   at the left
below corner of this picture. Indeed, according to  the
estimations which may be produced from (\ref{m1}),(\ref{m2} )
using mass of the most massive  fermion SM - the mass of
top-quark, we have $$m_{t-quark} =173,34 \pm 0.76 GeV$$ and for
Maximal Mass is equal to Planck's mass  $$M=10^{19} GeV $$ we can
obtain \be     m_1 = m_{t-quark}; \,\, m_2 ={m_{t-quark}}^2/2M
=1.9 \cdot10^{-15}GeV.\ee  And hence for non-Hermitian t-quark one
can see \be {m}_{t-quark}= m_{t-quark}+\gamma_5\cdot 2\cdot
10^{-6}{eV}.\ee Those we can see that non-Hermitian contribution
in this case is extremely small and practically non-observed. In
the same time the formulas (\ref{m1}),(\ref{m2}) for the lower
signs,  i.e. for exotic particles,  give values: \be
\tilde{m}_{t-quark} = 2M [1-{m^2}_{t-quark}/8M^2] +\gamma_5 2M
[1-{m^2}_{t-quark}/4M^2]. \ee

This means that all information about non-Hermitian exotic
fermions now concentrates on upper left corner of the
Fig.(\ref{Fig2}). And one can wait that most interesting phenomena
may be happening with the \emph{the lightest fermions.} Since
region of biggest mass $ m \sim M$ is still long  will be remain
unachieved  we may assume that this situation   resembles the
transition from \emph{non-relativistic to the relativistic
theory}. As it is known this transition was connected first of all
with experimental observations of \emph{limit values of the
speed}, which was associated with speed of light. Although the
light was always, but to measuring of this value it took  much
time!

Indeed, if the similar scenario according to the observation of
limited value of mass in mass-spectrum of elementary particles
will be produced, one should build accelerators till energies
$10^{19}$ GeV and this we must wait even much longer. However on
our happiness there is another possibility: searches consequences
of the existence of the Maximal Mass value in mass spectrum of
elementary particles based on the indirect experimental facts. As
the most important in this respect we can specify on the theory of
exotic particles, which is developed by us with the help of
non-Hermitian (pseudo-Hermitian) approach connected with
restrictions of mass spectrum of fermions. In this connection we
can say that different regions of undisturbed ${\cal PTS}$
ordinary and exotic particles are represented at Fig.(\ref{Fig3}).
It is easy to see that regions of undisturbed   ${\cal PT}$
symmetry for these  particles are strongly differ.

Performing calculations  in many ways reminiscent of similar
calculations carried out in the ordinary model in the magnetic
field \cite{TKR}, in a result, for modified Dirac-Pauli equations
one can find \emph{the exact solution for energy spectrum}
\cite{ROD1},\cite{ROD2}:
 \be\label{E61} E(\zeta,p_3,p_\perp
,\Delta\mu
H)=\sqrt{{p_3}^2-{m_2}^2+\left[\sqrt{{m_1}^2+{p_\perp}^2}+\zeta\Delta\mu
H \right]^2} \ee and for eigenvalues of the operator polarization
$\mu_3$ we can write in the form \be k=\sqrt{{m_1}^2
+{p_\perp}^2}, \ee where $p_3$ and $p_\perp$ are longitudinal and
transverse components of the  neutrinos momentum.

Dependence of of normalized values $E(-1,0, p_\bot, \Delta\mu H =
0.1 )/m$ on the parameter $x=m/M$ for $p_3$ and the cases
$p\bot=0,1,2,3,4$ and $\Delta\mu H = 0.1.$ can see at
Fig.(\ref{Fig8}).

From (\ref{E61}) it follows that in the field  where  ${\cal PT}$
symmetry is unbroken $m \leq M$, all energy levels are real for
the case of spin orientation along the magnetic field direction
$\zeta =+1$ (see Fig.(\ref{Fig1p1})).

However, in the opposite case $\zeta = -1$ (fermion spin is
oriented against the magnetic field) we have  appearance of  the
imaginary part energy from the ground state of fermion
$p_{\bot}=0$ and other low energy levels, see on
Fig.(\ref{Fig5-1}). There are also dependence of $E(-1,0, p_\bot,
\Delta\mu H = 0.1 )/M$ on the parameter $x=m/M$ for the different
values of transverse momentum components $p_\bot/m$:
$p\bot=0,1,2,3,4$ and $ \Delta\mu H = 0.1.$  In this case we can
see that regions of undisturbed symmetry in the linear
approximation of the intensity of the magnetic field $H$ are
defined by the following way \be H \leq \frac{{p_0}^2}{2 \Delta
\mu \sqrt{{m_1}^2 + {p_\bot}^2}},\ee where $p_0 = \sqrt{m^2
+{p\bot}^2 +{p_3}^2}$ - is the ordinary Hermitian energy of the
particles.

On the other hand dependence of $E(-1,p_3, p_\bot, \Delta\mu H =
0.1 )/M$ on the longitudinal components of momentum $x=p_3/M$ for
the different values of transverse momentum components $p_\bot/m$:
$p\bot=0,1,2,3,4$ and $\Delta\mu H = 0.1.$ is represented at
Fig.(\ref{Fig7}). The curves of type of dash are corresponding to
the anti-particles. In the case of fermion with spin orientated
along the magnetic field $\zeta = +1$ energetic states of
particles and anti-particles are  separated with the energetic
chink  order of $2 m_\nu$. But in the opposite case $\zeta = -1$
we can see that under the interaction of AMM with magnetic field
the different  branches of particles and anti-particles may be
intersected.

It is easy to see that in the case $ \Delta\mu =0$ from
(\ref{E61}) one can obtain the ordinary expression for energy of
charged particle in the magnetic field \emph{Landau levels}. For
this need to note that formula (\ref{E61})  is a valid not only
for the neutral fermions, but and for  charged particles
possessing AMM.  In this case one must simply replace the value of
transverse momentum of a neutral particle  on the quantum value of
charged fermions  in the magnetic field  ${p_\perp}^2\rightarrow
2\gamma n$, where  $\gamma=e H$ and $n=0,1,2...$ modified Landau
levels:

 \be\label{E621} E(\zeta,p_3,2\gamma n
,\Delta\mu H)=\sqrt{{p_3}^2-{m_2}^2+\left[\sqrt{{m_1}^2+{2\gamma
n}}+\zeta\Delta\mu H \right]^2} \ee
 Besides it should be emphasized that from the expression
(\ref{E621}), in the Hermitian limit putting $m_2=0$ and $m_1=m$
one can obtain expressions which  was early investigated in
\cite{TBZ}.

\section{ Discussions and conclusions}

It is well known \cite{n3},\cite{n33} that in the minimally
extended SM the one-loop radiative correction generates neutrino
magnetic moment which is proportional to the neutrino mass
\be\label{mu1}
  \mu_{\nu_e}=\frac{3}{8\sqrt{2}\pi^2}|e| G_F
  m_{\nu_e}=\left(3\cdot10^{-19}\right)\mu_0\left(\frac{m_{\nu_e}}{1
  eV}\right),
\ee where $ G_F$-Fermi coupling constant and $\mu_0$ is Bohr
magneton. Besides the discussion of problem of measuring the mass
of neutrinos (either active or sterile) show that for an active
neutrino model we have $\sum m_\nu =0.320 eV$, whereas for a
sterile neutrino $\sum m_\nu =0.06 eV$ \cite{n2}. However note
that the best laboratory upper limit on a neutrino magnetic
moment, $\mu\leq 2.9 10^{-11}\mu_0$, has been obtained by the
GEMMA collaboration \cite{mu1}, and the best astrophysical limit
is $\mu\leq 3\cdot10^{-12} \mu_0$.

The main feature of this approach is that even under the most
favorable conditions in respect to the values of the magnetic
moments which allow estimate the energy of ordinary neutrinos we
must to assume that corresponding intensity of the magnetic fields
simply fantastic huge. However for the case of exotic neutrinos we
have enormous enhancement of all effects proportionally to the
ratio of the Maximal Mass and the neutrinos mass of
$k=M/\tilde{m_\nu}$. And because the value of the Maximal Mass $M$
in the SM extends to infinity, the checking of presence of this
parameter is advisable start from the upper side of the possible
values. For example suitable for this purpose is the value of
Planck mass $M=10^{19}$ GeV. For these parameters the intensity of
magnetic fields may be  at the level of experimental achievements
of contemporary technology even for case of estimation of magnetic
moment in according to (\ref{mu1}).

Thus using (\ref{E61}) we can write for neutrino energy square \be
E^2(H)=p^2 +{\tilde{m}_\nu}^2 +2\zeta\Delta\mu
H\sqrt{m^2_1+{p_\bot}^2}+ (\Delta\mu H)^2.\ee
 If we consider the
case of cold neutrinos, momentum of  which is equal to zero, then
one could write for neutrino mass square in the magnetic field:

\be {\tilde{m}^2_\nu}(H)= {\tilde{m}_\nu}^2  \pm 4 \Delta\mu H
M.\ee The estimation of parameters $\Delta\mu H M $ with taking
into account that  Maximal Mass be equal to the Planck's mass and
value of magnetic moment value is determined by (\ref{mu1})  may
be represented in the form: \be \label{est} \Delta\mu H M =
10^{-19}\cdot\frac{e}{2m_e} H \cdot 10^{28}\cdot{1eV}^2
=\frac{10^{15}}{4}\left(\frac{H}{H_c}\right){1eV}^2,\ee where used
that $m_e$ is electron mass and $H_c$  \emph{characterizes quantum
magnetic field of electron} $H_c =4.41\cdot10^{13}$ Gauss. Note
that field of this strength will be able produce work on the
Compton wavelength of electron which  be equal to rest energy of
the electron. But from (\ref{mu1}) for the case of exotic
neutrinos we can see that  even the values of laboratory magnetic
intensity may produced a considerable effects.

 One can see that the main progress, is obtained by us in
the algebraic way of the construction of the fermion model with
$\gamma_5$-mass term is consists of describing of the new
energetic scale, which is defined by the parameter
$M={m_1}^2/2m_2$. This value on the scale of the masses is a point
of transition from the ordinary particles $m_2 < M$ to exotic
particle $m_2 > M $.

 At Fig.(\ref{Fig1111}) we can see the dependence of squared   mass of
  neutrinos    on  values   longitudinal     momentum $x=p_3$
  under fixed  values of its transverse components $p\bot$.
  Dependence of $E(-1,0, p_\bot, \Delta\mu H = 0.1 )/m$ on
the parameter $x=m/M$ for the different values of transverse
momentum components $p_\bot/m$: $p\bot=0,1,2,3,4\,\,
and\,\,\Delta\mu H = 0.1.$

As it was noted, if we suggest that exotic neutrinos have mass
$(\tilde{m}_\nu)$, which equals to mass of ordinary neutrinos then
perhaps, for this type of neutrinos also exist and extremely small
anomalous magnetic moments. But thanks to the huge value of
Maximal Mass comparable with Planck mass $M\approx 10^{19} GeV$
their interaction with magnetic field may be considerably enhanced
($K=M/\tilde{m}_\nu$). Hence experiments on search of exotic
neutrinos can simultaneously answer the question about the
existence of the limited spectrum of masses of elementary
particles.Therefore if somebody will be able to discover the
existence of the exotic neutrinos then it will mean that our World
is pseudo-Hermitian Universe and that it has a restriction of mass
spectrum of elementary particles.

As it is well known that experiments based  tritium  to measure
the absolute   value  mass of the neutrino have a long history.
The issue of atomic and molecular   excitations in tritium, based
on neutrino experiments was first raised in the early 1970s
\cite{12}.   There   was able in the  first time to set a limit of
neutrinos mass about 55-eV  \cite{13} and has arisen an
understanding about more lower limits further.  The Los Alamos
National Laboratory (LANL)-experiment produced an upper limit of
${m_\nu} < 9.3 eV$  at the 95\% confidence level \cite{15} was
obtained with a $2\sigma$  excess. These  events was  observed in
the endpoint of energetic region and  quantitatively reported  at
first time about \emph{a negative values of central  points of}
${m_{\nu}}^2$. An experiment at Lawrence Livermore National
Laboratory (LLNL),  also using  a  gaseous $T_2$ source, has
fixed, which a central value is in good agreement  with the LANL
result,  but with much reduced statistical uncertainties.

 In  concurrent experiments \cite{17}- \cite{19} has used complex
  tritium sources.  However all of these experiments gave results that
  were consistent with  zero neutrino mass \emph{but with  the central  values
  are   in regions of the  negative-mass squared.} These values were symptomatic
  and witnessed about  an insufficiently successful coordination of
  theoretical and experimental  results for explanation nagative  dependence
  of mass squared \cite{20}. Attempts to reduce diapason of such conflicting
  results  furthered interest in molecular-tritium experiments.  The  limit on
  the neutrino mass,$m < 2 eV $ at an unstated confidence level,  is derived
  from the Mainz \cite{21} and Troitsk \cite{22}, \cite{23} experiments, both of
  which employed a new type  of spectrometers. In these experiments a
  magnetic-adiabatic collimation-with-electrostatic (MAC-E) filter \cite{24}
  were  used.  Beta electrons were rotated  are in a \emph{region of large magnetic field.}

The Troitsky's experiment,  like its predecessors at LANL and
LLNL, used a   windowless, gaseous tritium source.  The gas
density and source purity   were monitored indirectly by a mass
analyzer at the  source and by   count-rate measurements at a low
potential setting. After that  an  electron gun was run, which
mounted upstream of the source.  The initial analysis of the data
required the inclusion  of a step function which  added to the
spectral shape \cite{22},   so-called "Troitsk  anomaly."
 This result, based  on a re-analysis of the  source  calibrations was presented
 in the form $m < 2.05 eV$ at 95\% confidence \cite{23}. But  at last  the final
 results of Troitsky's  experiment,  which produces   negative square mass  in
 central point of energetic region of neutrinos, can  be represented in the form  \cite{nu}:

\be\label{100} {m_\nu}^2 = -0.67 \pm1.89_{stat} \pm 1.68_{syst}
{eV}^2,\ee at 95\% confidence(C.L).

 And although in this paper it is noted,
that " the result for the mass of neutrino is negative, but the
deviation from zero is not statistically significantly". However
authors of article \cite{nu} also noted that in one of  sessions
measurement negative values of square  mass neutrinos was obtained
the result even \be {m_\nu}^2 = - 8.06 \pm 6.99 \pm 1.65 {eV}^2
.\ee In this situation we want draw attention to the fact that one
can assume we simply deal with experiment measurements of
neutrinos mass \emph{with taking into account influence of the
magnetic fields.} According to scheme of the the Troitsk
experimental installation for measurement of mass the electron
antineutrino gas-tritium is injected into a long tube, where  a
strong magnetic field   (up to $H = 0.8 T = 8000 Gauss$) is
existed. If we suggest that the beam of neutrinos partially
consist   of pseudo-Hermitian components exotic neutrinos($
\tilde{m} $) that we can obtain the following evaluation

 \be\label{m2nu}
\tilde{\tilde{m}_\nu}^2(H)=\tilde{m_\nu}^2 - 4 \Delta\mu H M. \ee
Indeed if we consider values, which followed from  the ordinary
experimental data and using $m_\nu= 1 eV$ and formulas
(\ref{100}),(\ref{m2nu}) with ${\tilde{m}_\nu}^2(H) = -1 {eV}^2$
one can write \be\label{DeltaH} 4\Delta\mu H M \simeq 2 {eV}^2.\ee
Using the last expressions we obtain that observing results
follows if the strength of the magnetic fields $H \simeq 8000
Gauss$ \cite{nu} and Maximal mass value is equal to $2\cdot10^{14}
GeV$ when $\mu_\nu $ taken from (\ref{mu1}).  However if the
values of magnetic moment of neutrinos grows in comparison with
(\ref{mu1}) then level of the restriction Maximal Mass can be
reduced considerably.

\end{document}